\documentstyle[osa,opex]{revtex}
\begin{document}
 
\title{Maxwell's Demon at work: \\
    Two types of Bose condensate fluctuations in power-law traps}

\author{Siegfried Grossmann and Martin Holthaus}
\address{Fachbereich Physik der Philipps-Universit\"at,
    Renthof 6, D-35032 Marburg, Germany}
\email{holthaus@stat.physik.uni-marburg.de}

\begin{abstract}
After discussing the key idea underlying the Maxwell's Demon ensemble,
we employ this idea for calculating fluctuations of ideal Bose gas
condensates in traps with power-law single-particle energy spectra.
Two essentially different cases have to be distinguished. If the heat
capacity remains continuous at the condensation point in the
large-$N$-limit, the fluctuations of the number of condensate particles
vanish linearly with temperature, independent of the trap characteristics.
If the heat capacity becomes discontinuous, the fluctuations vanish
algebraically with temperature, with an exponent determined by the trap.
Our results are based on an integral representation that yields the solution
to both the canonical and the microcanonical fluctuation problem in a
singularly transparent manner.   
\end{abstract}

\begin{OEReferences}
\item\label{LandauLifshitz59} L.D. Landau and E.M. Lifshitz,
    {\em Statistical Physics\/}
    (Pergamon, London, 1959).
\item\label{Pathria85} R.K. Pathria,
    {\em Statistical Mechanics\/}
    (Pergamon, Oxford, 1985).
\item\label{FujiwaraEtAl70} I. Fujiwara, D. ter Haar, and H. Wergeland,
    ``Fluctuations in the population of the ground state of Bose systems'',
    {\em J.\ Stat.\ Phys.} {\bf 2}, 329-346 (1970).
\item\label{ZiffEtAl77} R.M. Ziff, G.E. Uhlenbeck, and M. Kac,
    ``The ideal Bose--Einstein gas, revisited'',
    {\em Phys.\ Rep.} {\bf 32}, 169-248 (1977).
\item\label{GajdaRzazewski97} M. Gajda and K. Rz\c{a}\.{z}ewski,
    ``Fluctuations of Bose--Einstein condensate'',
    {\em Phys.\ Rev.\ Lett.} {\bf 78}, 2686-2689 (1997).	
\item\label{AndersonEtAl95} M.H. Anderson, J.R. Ensher, M.R. Matthews,
    C.E. Wieman, and E.A. Cornell,
    ``Observation of Bose--Einstein condensation in a dilute atomic vapor'',
    {\em Science} {\bf 269}, 198-201 (1995).
\item\label{DavisEtAl95} K.B. Davis, M.-O. Mewes, M.R. Andrews,
    N.J. van Druten, D.S. Durfee, D.M. Kurn, and W. Ketterle,
    ``Bose--Einstein condensation in a gas of sodium atoms'',
    {\em Phys.\ Rev.\ Lett.} {\bf 75}, 3969-3973 (1995).		
\item\label{BradleyEtAl97} C.C. Bradley, C.A. Sackett, and R.G. Hulet,
    ``Bose--Einstein condensation of lithium: observation of limited
    condensate number'',
    {\em Phys.\ Rev.\ Lett.} {\bf 78}, 985-989 (1997).
\item\label{KetterleDruten96} W. Ketterle and N.J. van Druten,
    ``Bose--Einstein condensation of a finite number of particles
    trapped in one or three dimensions'',
    {\em Phys.\ Rev.\ A} {\bf 54}, 656-660 (1996).
\item\label{DrutenKetterle97} N.J. van Druten and W. Ketterle,
    ``Two-step condensation of the ideal Bose gas in highly
    anisotropic traps'',
    {\em Phys.\ Rev.\ Lett.} {\bf 79}, 549-552 (1997).
\item\label{GH96} S. Grossmann and M. Holthaus,
    ``Microcanonical fluctuations of a Bose system's ground state
    occupation number'',
    {\em Phys.\ Rev.\ E} {\bf 54}, 3495-3498 (1996).
\item\label{GH97a} S. Grossmann and M. Holthaus,
    ``From number theory to statistical mechanics: Bose--Einstein
    condensation in isolated traps'',
    to appear in {\em Chaos, Solitons \& Fractals\/} (Proceedings
    of the 178th Heraeus-Seminar {\em Pattern formation in nonlinear
    optical systems\/}, Bad Honnef, June 23--25, 1997).
\item\label{Wilkens96} M. Wilkens,
    ``From Chinese wok to Mexican hat: Bose--Einstein condensation
    in an isolated Bose gas''
    (Preprint, Konstanz, 1996).
\item\label{NavezEtAl97} P. Navez, D. Bitouk, M. Gajda, Z. Idziaszek,
    and K. Rz\c{a}\.zewski,
    ``The fourth statistical ensemble for the Bose--Einstein condensate'',
    {\em Phys.\ Rev.\ Lett.} {\bf 79}, 1789-1792 (1997).
\item\label{GH97b} S. Grossmann and M. Holthaus,
    ``Fluctuations of the particle number in a trapped Bose--Einstein
    condensate'',
    {\em Phys.\ Rev.\ Lett.} {\bf 79}, 3557-3560 (1997).	     
\item\label{BorrmannFranke93} P. Borrmann and G. Franke,
    ``Recursion formulas for quantum statistical partition functions'',
    {\em J.\ Chem.\ Phys.} {\bf 98}, 2484-2485 (1993).
\item\label{Eckhardt97} B. Eckhardt,
    ``Eigenvalue statistics in quantum ideal gases''.
    In: {\em Emerging applications of number theory\/},
    edited by D. Hejhal, F. Chung, J. Friedman, M. Gutzwiller,
    and A. Odlyzko
    (Springer, New York, to appear 1997).
\item\label{WilkensWeiss97} M. Wilkens and C. Weiss,
    ``Universality classes and particle number fluctuations of trapped
    ideal Bose gases''
    (Preprint, Potsdam, 1997).
\item\label{deGrootEtAl50} S.R. de Groot, G.J. Hooyman, and C.A. ten Seldam,
    ``On the Bose--Einstein condensation'',
    {\em Proc.\ Roy.\ Soc.\ London A} {\bf 203}, 266-286 (1950).
\item\label{Nanda53} V.S. Nanda,
    ``Bose--Einstein condensation and the partition theory of numbers'',
    {\em Proc.\ Nat.\ Inst.\ Sci.\ (India)} {\bf19}, 681-690 (1953).
\item\label{Politzer96} H.D. Politzer,
    ``Condensate fluctuations of a trapped, ideal Bose gas'',
    {\em Phys.\ Rev.\ A} {\bf 54}, 5048-5054 (1996).
\item\label{WuEtAl96} F.Y. Wu, G. Rollet, H.Y. Huang, J.M. Maillard,
    C.-K. Hu, and C.-N. Chen,
    ``Directed compact lattice animals, restricted partitions of an
    integer, and the infinite-states Potts model'',
    {\em Phys.\ Rev.\ Lett.} {\bf 76}, 173-176 (1996).		
\end{OEReferences}

\noindent
According to grand canonical statistics, the root-mean-square fluctuations
$\delta N_\nu$ of the occupation numbers $N_\nu$ of an ideal Bose gases's
$\nu$-th single-particle state are given
by~[\ref{LandauLifshitz59},\ref{Pathria85}]
\begin{equation}
\left(\delta N_\nu\right)^2 = N_\nu(N_\nu + 1)	\; . 
\label{FLUC}
\end{equation}
This expression follows without any approximation from the grand canonical
approach, but it faces a severe problem when applied to the fluctuations
$\delta N_0$ of the ground state occupation number $N_0$ of {\em isolated\/}
Bose gases: if the temperature approaches zero, all $N$ particles of an
isolated Bose gas occupy the ground state, so that the actual fluctuations
vanish, whereas Eq.~(\ref{FLUC}) predicts fluctuations $\delta N_0$ of the
order $N$. This seems to be one of the most important examples where the
different statistical ensembles can not be regarded as equivalent. When
computing low-temperature fluctuations of the ground state occupation number
for isolated Bose gases, one therefore has to give up the convenient grand
canonical point of view, and to resort to a microcanonical treatment.

Although this problem had been well recognized and discussed some time
ago~[\ref{FujiwaraEtAl70},\ref{ZiffEtAl77}], tools for computing the
microcanonical fluctuations $\delta N_0$ have been developed only
recently~[\ref{GajdaRzazewski97}], spurred by the progress in preparing
Bose--Einstein condensates of alkali atoms in magnetic
traps~[\ref{AndersonEtAl95},\ref{DavisEtAl95},\ref{BradleyEtAl97}].
A particularly instructive model system for illustrating the microcanonical
approach to fluctuations $\delta N_0$ is provided by $N$ ideal Bosons
trapped in a one-dimensional harmonic potential. Since quasi one-dimensional
harmonic trapping potentials can be realized as limiting cases of strongly
anisotropic three-dimensional
traps~[\ref{KetterleDruten96},\ref{DrutenKetterle97}],
this system is not merely of academic interest. The value of the model
lies in the fact that it allows one to map the problem of evaluating the
microcanonical statistics to problems also arising in analytic number theory,
because the number of microstates accessible at some excitation energy~$E$
equals the number of partitions of the integer $n = E/(\hbar\omega)$ into
no more than $N$ summands, with $\omega$ being the oscillator frequency.
Using the appropriate asymptotic formulae from partition theory, one finds
that the microcanonical fluctuations $\delta N_0$ for this model system
vanish linearly with temperature~$T$~[\ref{GH96},\ref{GH97a}]:        
\begin{equation}
\delta N_0 \; \approx \; \frac{\pi}{\sqrt 6} \frac{k_B T}{\hbar\omega}
    \qquad {\mbox{for}} \qquad
    T \ll T_0^{(1)} \equiv \frac{\hbar\omega}{k_B} \frac{N}{\ln N} \; ,
\label{ODOF}
\end{equation}
where $k_B$ is the Boltzmann constant, and $T_0^{(1)}$ denotes the
temperature below which the ground state occupation becomes significant.
As illustrated in Fig.~1, which compares the relative microcanonical
fluctuations $\delta N_0/N$ to the corresponding grand canonical
fluctuations and to the approximation~(\ref{ODOF}), for $N = 10^6$
particles, this approximation is quite good indeed. The very same
result~(\ref{ODOF}) has also been obtained by Wilkens~[\ref{Wilkens96}]
within a {\em canonical\/} approach, that is, for a trap in contact with
a heat bath. 

\vspace*{0.3cm}
\noindent
\centerline{\epsfxsize=3in \epsfbox{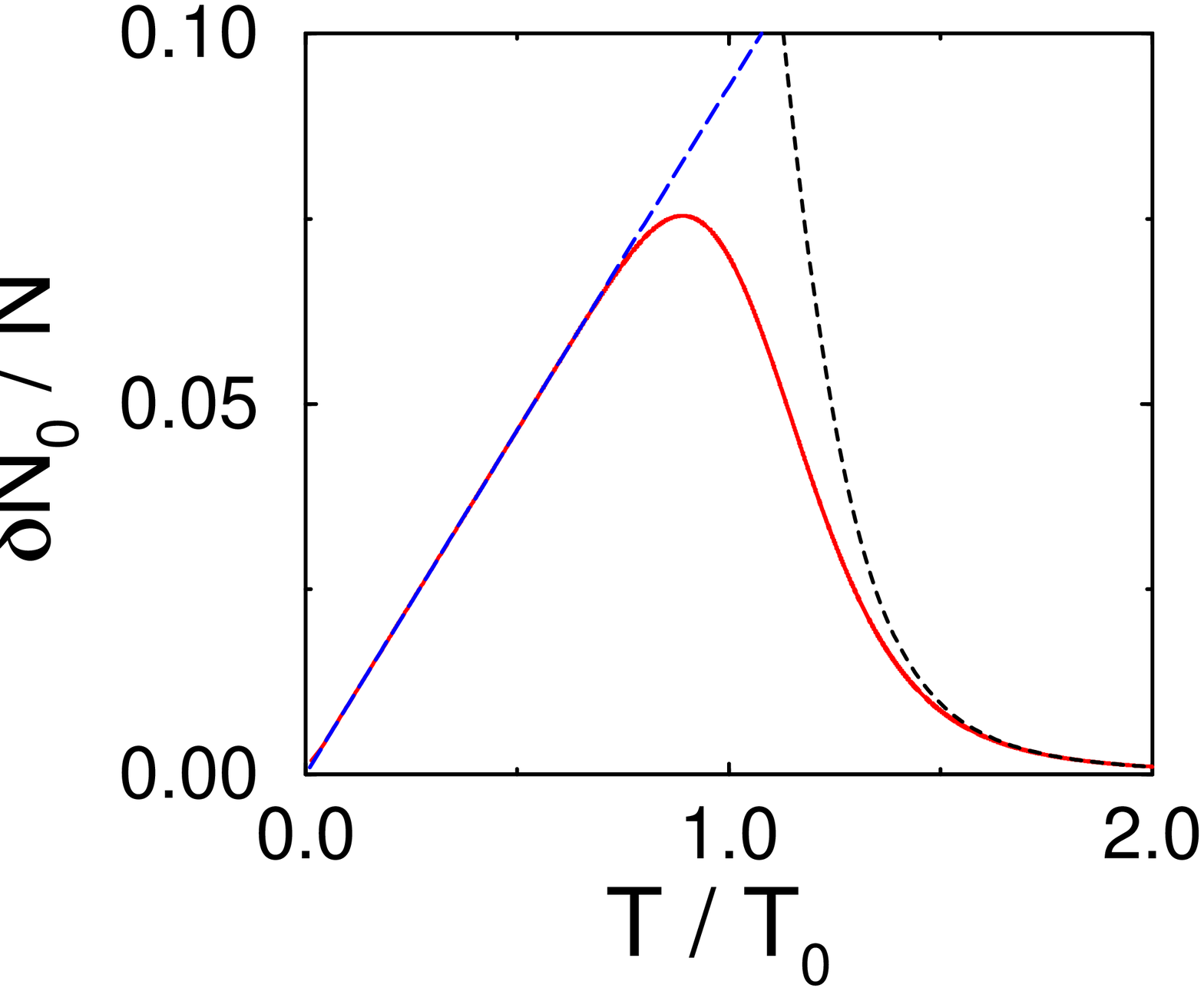}}

\vspace*{-0.2cm}
\noindent
{\footnotesize Fig.~1. Full red line: microcanonical fluctuations
    $\delta N_0/N$ for a system of $N = 10^6$ ideal Bose particles
    trapped by a one-dimensional harmonic potential~[\ref{GH96}]. The
    temperature $T_0 \equiv T_0^{(1)}$ denotes the characteristic
    temperature below which the ground state occupation becomes
    significant, see Eq.~(\ref{ODOF}).
    Black short-dashed line: grand canonical fluctuations for the same
    system.
    Blue dashed line: low-temperature approximation provided by
    Eq.~(\ref{ODOF}).}
\vspace*{5mm}

How can one generalize this finding to other trap types? A rather interesting
suggestion has been made by Navez {\em et al.\/}~[\ref{NavezEtAl97}].
Denoting, for an ideal $N$-particle Bose gas in some arbitrary trap, the
number of microstates with {\em exactly\/} $N_{\rm ex}$ excited particles as
$\Phi(N_{\rm ex}|E)$, so that the total number of microstates accessible
at the given energy $E$ reads
$\Omega(E|N) = \sum_{N_{\rm ex}=0}^{N} \Phi(N_{\rm ex}|E)$, 
these authors consider the generating function
\begin{equation}
\Upsilon(z,E) \; = \sum_{N_{\rm ex}=0}^{\infty} \!
    z^{N_{\rm ex}} \, \Phi(N_{\rm ex}|E)	\; .
\label{MWDF}
\end{equation}
This function involves $\Phi(N_{\rm ex}|E)$ even for $N_{\rm ex} > N$,
which appears to be unphysical: after all, the excitation energy $E$ can not
be distributed over more than the~$N$ particles. However, {\em provided\/}
the microcanonical distributions for finding $N_{\rm ex}$ out of $N$
particles in an excited trap state, 
\begin{equation}
p_{\rm ex}(N_{\rm ex}|E) \; = \;  
    \frac{\Phi(N_{\rm ex}|E)}{\Omega(E|N)} \; ,
    \qquad N_{\rm ex} = 1,2,\ldots,N \; ,
\label{DIST}
\end{equation}
are strongly peaked around some value $\overline{N}_{\rm ex} \ll N$,
which should be the case for temperatures well below the onset of
Bose--Einstein condensation, we will have
$\Phi(N_{\rm ex}|E)/\Omega(E|N) \approx 0$ for $N_{\rm ex} > N$. In that
case the generating function~(\ref{MWDF}) would be quite useful, since one
could obtain the microcanonical expectation value $\langle N_0 \rangle$ for
the ground state occupation number, and its fluctuation, from
\[
N - \langle N_0 \rangle \; = \;
    \left. z\frac{\partial}{\partial z}
    \ln \Upsilon(z,E) \right|_{z=1}
\qquad {\mbox{and}} \qquad
\left(\delta N_0\right)^2 \; = \;
    \left. \left(z\frac{\partial}{\partial z}\right)^2
    \ln \Upsilon(z,E) \right|_{z=1} \; , 
\]
respectively~[\ref{NavezEtAl97}]. The proviso can be formulated in more
intuitive terms: the required well-peakedness of the distributions~(\ref{DIST})
means that those microstates where the energy $E$ is actually spread out
over all $N$ particles carry only negligible statistical weight, so that
the overwhelming majority of all microstates leaves a fraction of the
particles in the ground state, forming the Bose condensate. Then the
restriction on the number of microstates caused by the fact that there
is only a finite number~$N$ of particles becomes meaningless, so that,
loosely speaking, ``the system has no chance to know how many particles
the condensate consists of''. But if this is the case, i.e., if the system's
properties become insensitive to the actual number of particles contained
in the condensate, then one can act as if the condensate particles constituted
an {\em infinite\/} reservoir. Thus, the generating function~(\ref{MWDF}) may
be regarded as the partition function of a rather unusual ensemble, consisting
of the excited-states subsystems of Bose gases that exchange particles with
the ground state ``reservoirs'' without exchanging energy. Since such an
exchange process, if performed by hand, requires a genius who is able
to separate the hot, excited particles from the cold ones in the ground
state, this new ensemble has been called the ``Maxwell's Demon
ensemble''~[\ref{NavezEtAl97}].

But can we rely on Maxwell's Demon, that is, does the proviso hold?
This question needs to be answered first. A strong argument in favour of
the Maxwell's Demon ensemble has already been provided by the
approximation (\ref{ODOF}) to the low-temperature fluctuations for a Bose
gas in a one-dimensional oscillator potential: these fluctuations are
{\em independent of the total particle number\/} $N$, as they should be
if the system really has no knowledge of the number of condensate
particles, and thus of $N$. The awe-inspiring agreement with the actual
microcanonical fluctuations depicted in Fig.~1 leaves no doubt
that this approximation {\em is\/} reliable. To further substantiate the
new ensemble, we also consider the microcanonical fluctuations $\delta N_0$
for an ideal Bose gas trapped by a three-dimensional isotropic harmonic
oscillator potential~[\ref{GH97b}]. The numbers $\Omega(E|N)$ of
microstates for some given excitation energy $E = n\hbar\omega$ can then be
obtained from the canonical $N$-particle partition function
\begin{equation}
Z_N(\beta) \; = \; \sum_{n=0}^{\infty} e^{-n\beta\hbar\omega} \,
    \Omega(n\hbar\omega|N) \; ,
\label{CNPF}  
\end{equation}
which, in turn, can be calculated numerically with the help of the recursion
relation~[\ref{BorrmannFranke93},\ref{Eckhardt97},\ref{WilkensWeiss97}]
\begin{equation}
Z_N(\beta) \; = \; \frac{1}{N}\sum_{k=1}^{N}
    Z_1(k\beta) Z_{N-k}(\beta)	\; .
\end{equation}
As usual, $\beta = 1/(k_BT)$ denotes the inverse temperature. 
By means of numerical saddle-point inversions of Eq.~(\ref{CNPF}), we
compute the desired numbers $\Omega(E|N_{\rm ex})$ for $N_{\rm ex}$
ranging from $1$ to $N$~[\ref{GH97b}], and get the differences
\begin{equation}
\Phi(N_{\rm ex}|E) \; = \; \Omega(E|N_{\rm ex}) - \Omega(E|N_{\rm ex}-1)
\label{MDIF}
\end{equation}
that determine the microcanonical distributions~(\ref{DIST}). Some of
these distributions are displayed in Fig.~2, for $N = 1000$ and several
``low'' temperatures. What we find is exactly what is needed for Maxwell's
Demon: the distributions are well peaked for temperatures below the onset
of condensation, and remarkably close to Gaussians~[\ref{GH97a}]. It is then
no surprise that the corresponding microcanonical low-temperature fluctuations
$\delta N_0$, obtained from the widths of these distributions, are --- as
long as a condensate exists! --- once again independent of $N$, as
exemplified in Fig.~3 for $N = 200$, $500$, and $1000$. As discussed above,
it is precisely this $N$-independence, expressed mathematically by the
appearance of the upper summation bound ``$\infty$'' rather than ``$N$'' in
Eq.~(\ref{MWDF}), that lies at the bottom of the Maxwell's Demon ensemble.
But whereas this $N$-independence is, by construction, {\em put into\/}
this ensemble, it has {\em come out\/} here as the result of a truly
microcanonical calculation~[\ref{GH97a},\ref{GH97b}] that works with
the actual $N$, not with $\infty$. 

\vspace*{0.3cm}
\noindent
\centerline{\epsfxsize=3in \epsfbox{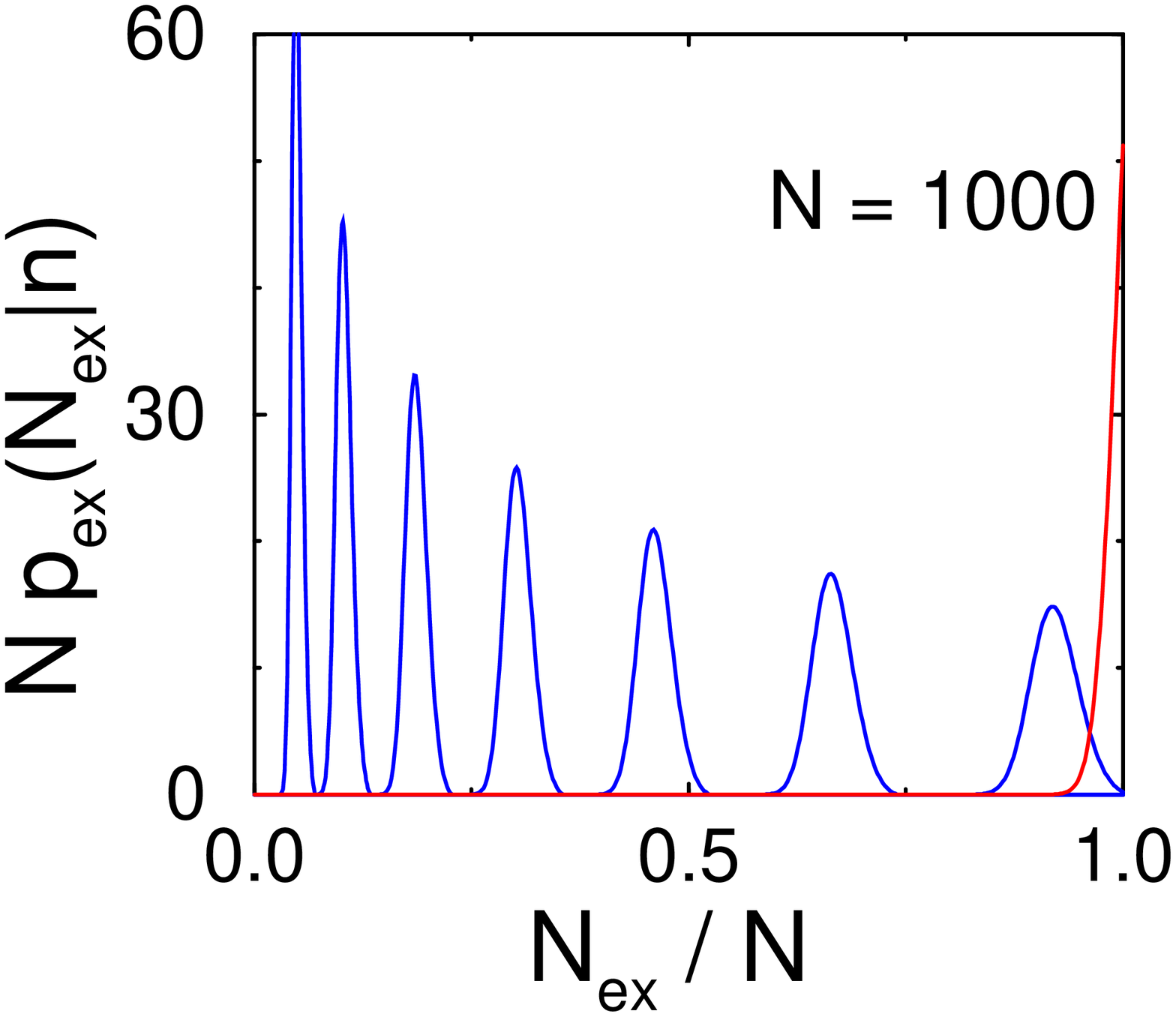}}

\vspace*{-0.2cm}
\noindent
{\footnotesize Fig.~2. Microcanonical probability distributions
    $p_{\rm ex}(N_{\rm ex}|n)$ for finding $N_{\rm ex}$ out of $N=1000$
    ideal Bose particles, trapped by a three-dimensional isotropic harmonic
    potential, excited when the total excitation energy $E$ is
    $n \cdot \hbar\omega$, with $\omega$ denoting the oscillator frequency.
    The number $n$ determines the temperature $T$.
    The normalized temperatures $T/T_0$ corresponding to the blue,
    Gaussian-like distributions range from $0.3$ to $0.9$ (left to right,
    in steps of $0.1$); $T_0 = (\hbar\omega/k_B)(N/\zeta(3))^{1/3}$.
    Due to finite-$N$-effects, the condensation temperature is lowered from
    $T_0$ to about $0.93 \, T_0$. The temperature corresponding to the
    rightmost, red distribution is $T = 0.95 \, T_0$, lying slightly
    above the condensation point.}
\vspace{2mm}

\vspace*{0.3cm}
\noindent
\centerline{\epsfxsize=3in \epsfbox{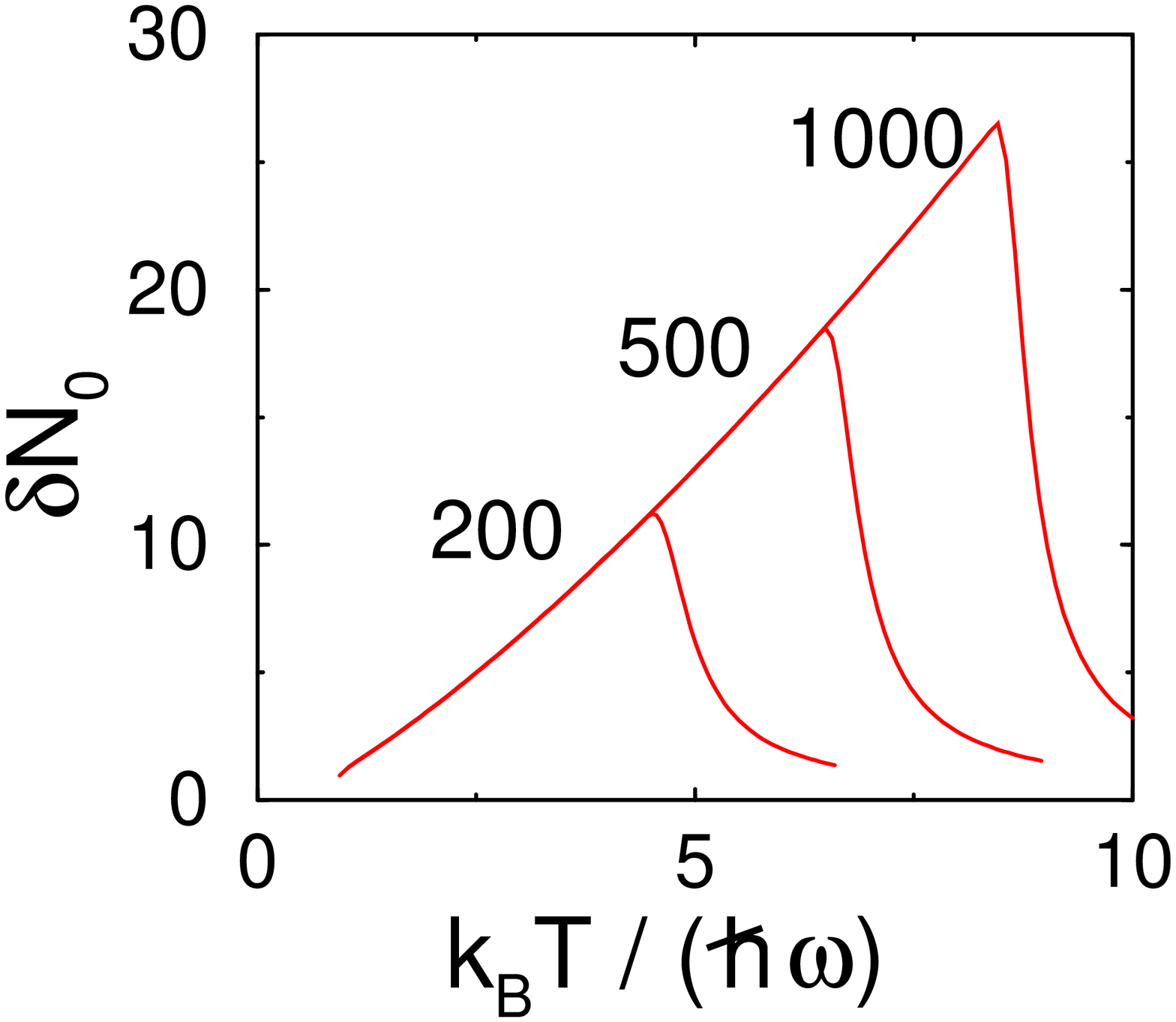}}

\vspace*{-0.2cm}
\noindent
{\footnotesize Fig.~3. Microcanonical fluctuations $\delta N_0$ for
    $N = 200$, $500$, and $1000$ ideal Bose particles trapped by a
    three-dimensional, isotropic harmonic potential. The fluctuations are
    maximal close to the respective condensation points. These maximal
    fluctuations scale approximately as $\sqrt{N}$, cf.~Eqs.~(\ref{CONT})
    and~(\ref{MFL1}). Note that the low-temperature fluctuations for
    all three systems agree perfectly, thus demonstrating the
    $N$-independence of $\delta N_0$ below the condensation point.}
\vspace{5mm}

Having thus gained confidence in the abilities of Maxwell's Demon, we now
set it to work in order to compute condensate fluctuations $\delta N_0$.
To this end, we consider ideal Bose gases in $d$-dimensional traps with
arbitrary single-particle energies ${\varepsilon_{\nu}}$; we stipulate
$\varepsilon_0 = 0$. Denoting the grand canonical partition function by
$\Xi(z,\beta)$, we base our analysis on its ``excited'' part
$\Xi_{\rm ex}(z,\beta) \equiv (1 - z)\Xi(z,\beta)$.
Since, by virtue of Eq.~(\ref{MDIF}), 
\begin{equation}
\Xi_{\rm ex}(z,\beta) \; = \sum_{N_{\rm ex}=0}^{\infty} \! z^{N_{\rm ex}} \!
    \sum_{E} \, \Phi(N_{\rm ex}|E) \, e^{-\beta E} \; ,
\end{equation}
this function has the decisive property 
\begin{equation}
\left.
\left(z\frac{\partial}{\partial z}\right)^{\!k} \!
    \Xi_{\rm ex}(z,\beta) \right|_{z=1}
    = \; \sum_E \left(
    \sum_{N_{\rm ex}=0}^{\infty} N_{\rm ex}^k \, \Phi(N_{\rm ex}|E)
    \right) e^{-\beta E} \; \equiv \; M_k(\beta) \; ,
\label{RICH}
\end{equation}
i.e., it yields directly the non-normalized {\em canonical\/} moments
$M_k(\beta)$, and generates the {\em microcanonical\/} moments
\begin{equation}
\mu_k(E) \; \equiv \;
    \sum_{N_{\rm ex}=0}^{\infty} \, N_{\rm ex}^k \, \Phi(N_{\rm ex}|E)
    \qquad {\mbox{with}} \qquad k = 0,1,2,\ldots \; .
\label{MIMO}
\end{equation} 
We then employ the Maxwell's Demon approximation: as long as there is a
condensate, these moments (with $N_{\rm ex}$ ranging from $0$ to $\infty$)
approximate the true moments of the physical set
$\{ \, \Phi(N_{\rm ex}|E) \, \}$
(where the number $N_{\rm ex}$ of excited particles can not exceed the
total particle number $N$); in this approximation one has the identity 
$\mu_0(E) = \Omega(E|N)$~[\ref{GH97b}]. Now the calculations within the
canonical ensemble become remarkably simple. The canonical expectation
value $\langle N_{\rm ex} \rangle = N - \langle N_0 \rangle$
of the number of excited particles is given by
\begin{equation}
\langle N_{\rm ex} \rangle \; = \; \frac{M_1(\beta)}{M_0(\beta)}    \; ;
\end{equation}
the canonical condensate fluctuations
$\left(\delta N_0\right)^2_{cn} \; = \; \left(\delta N_{\rm ex}\right)^2_{cn}$
follow from
\begin{equation}
\left(\delta N_0\right)^2_{cn} \; = \; \frac{M_2(\beta)}{M_0(\beta)}
    - \left( \frac{M_1(\beta)}{M_0(\beta)} \right)^2 \; .
\end{equation}
Without any further approximation, these expressions can be rewritten
as complex integrals:
\begin{equation}
\langle N_{\rm ex} \rangle \; = \;
    \frac{1}{2\pi i} \int_{\tau-i\infty}^{\tau+i\infty} \!\! {\mbox d}t \,
    \Gamma(t) \, Z(\beta,t) \, \zeta(t)
\label{FUN1}
\end{equation}
and
\begin{equation}
\left(\delta N_0\right)^2_{cn} \; = \; 
    \frac{1}{2\pi i} \int_{\tau-i\infty}^{\tau+i\infty} \!\! {\mbox d}t \,
    \Gamma(t) \, Z(\beta,t) \, \zeta(t-1) \; ,
\label{FUN2}
\end{equation}
where $\Gamma(t)$ and $\zeta(t)$ denote the Gamma function and
Riemann's Zeta function, respectively. All the information about
the specific trap under consideration is embodied in its spectral
Zeta function
\begin{equation}
Z(\beta,t) \; = \;
    \sum_{\nu=1}^{\infty} \frac{1}{(\beta\varepsilon_\nu)^t} \;
\end{equation}
where the sum runs over the trap spectrum, excluding the ground state
energy $\varepsilon_0$~$=$~$0$. The real number $\tau$ in Eqs.~(\ref{FUN1})
and (\ref{FUN2}) has to be chosen such that the path of integration up
the complex $t$-plane sees all poles to its left. 

So far, the analysis is quite general. We now specialize the further
deliberations to ideal Bose gases in $d$-dimensional traps with power-law
single-particle spectra
\begin{equation}
\varepsilon_{\{\nu_i\}} = \Delta \sum_{i=1}^{d} c_i \nu_i^\sigma  \; ,
    \qquad \nu_i = 0,1,2,\ldots \; , \qquad \sigma > 0 \; ,
\label{SPSP}
\end{equation}
where the dimensionless coefficients $c_i$ characterize the trap's
anisotropy, normalized such that the lowest $c_i$ is unity; the
characteristic energy $\Delta$ measures the gap between the ground state
and the first excited state, and the exponent $\sigma$ is determined by
the potential's shape. Such systems have been studied first by
de Groot {\em et al.\/}~[\ref{deGrootEtAl50}]; we have adopted here the
notation also employed by Wilkens and Weiss~[\ref{WilkensWeiss97}]. 

If we consider $N$-asymptotically large systems and disregard
finite-$N$-effects, that is, if we focus on gases consisting
of at least some $10^5$ particles, say, then a good approximation to
the density of states is provided by 
\begin{equation}
\rho(E) \; = \; \frac{A}{\Gamma\!\left(\frac{d}{\sigma}\right)}
    \left(\frac{E}{\Delta}\right)^{\!d/\sigma-1} \!\frac{1}{\Delta}
    \qquad {\mbox{with}} \qquad
A \; \equiv \; \frac{\Gamma\!\left(\frac{1}{\sigma}+1\right)^d}
    {\left(\prod_{i=1}^d c_i\right)^{\!1/\sigma}}  \; .
\end{equation}
Using this density, and assuming that the anisotropy coefficients $c_i$
are not too different from each other~[\ref{DrutenKetterle97}], the usual
line of reasoning shows that for $d/\sigma>1$ there is a sharp onset of
Bose--Einstein condensation at the temperature $T_0$ given
by~[\ref{deGrootEtAl50}] 
\begin{equation}
\frac{k_B T_0}{\Delta} \; = \; \frac{1}{A^{\sigma/d}}
    \left(\frac{N}{\zeta\!\left(\frac{d}{\sigma}\right)}\right)^{\!\sigma/d}
    \; .
\label{CONT}
\end{equation}
Moreover, the spectral Zeta functions can now be well approximated by
\begin{equation}
Z(\beta,t) \; \approx \; \frac{A}{\Gamma\!\left(\frac{d}{\sigma}\right)}
    \left(\beta\Delta\right)^{-t}
    \zeta(t+1-d/\sigma)	\; ,
\end{equation}
so Eqs.~(\ref{FUN1}) and (\ref{FUN2}) adopt the transparent forms 
\begin{equation}
\langle N_{\rm ex} \rangle \; \approx \;
    \frac{A}{\Gamma\!\left(\frac{d}{\sigma}\right)}
    \frac{1}{2\pi i} \int_{\tau-i\infty}^{\tau+i\infty} \!\! {\mbox d}t \,
    \left(\beta\Delta\right)^{-t} \Gamma(t) \,
    \zeta(t + 1 - d/\sigma) \, \zeta(t)
\label{GREX}
\end{equation}
and
\begin{equation}
\left(\delta N_0\right)^2_{cn} \; \approx \; 
    \frac{A}{\Gamma\!\left(\frac{d}{\sigma}\right)}
    \frac{1}{2\pi i} \int_{\tau-i\infty}^{\tau+i\infty} \!\! {\mbox d}t \,
    \left(\beta\Delta\right)^{-t} \Gamma(t) \,
    \zeta(t + 1 - d/\sigma) \, \zeta(t-1)	\; .
\label{GRFL}
\end{equation}

For $\Delta \ll k_BT$, the behavior of either integral is determined by the
pole of its integrand farthest to the right in the complex plane. Keeping in
mind that $\zeta(z)$ has merely one single pole at $z = 1$, with residue~$1$,
while the poles of $\Gamma(z)$ are located at $z = 0$, $-1$, $-2$, \ldots, 
the decisive pole is provided {\em either\/} by the system's Zeta function
$\zeta(t + 1 - d/\sigma)$, {\em or\/} by the other Zeta function that is
determined by the order of the cumulant one is asking for: by $\zeta(t)$, if
one asks for the first cumulant $\langle N_{\rm ex} \rangle$, or by
$\zeta(t-1)$, if one asks for the second cumulant $\left(\delta N_0\right)^2$.
To see what the argument boils down to, let us first consider the evaluation
of Eq.~(\ref{GREX}), where the system's pole at $t = d/\sigma$ competes with
the cumulant-order pole at $t=1$:  		
\begin{itemize}
\item If $d/\sigma > 1$, the low-temperature behavior of
$\langle N_{\rm ex} \rangle$ is governed by the pole of
$\zeta(t + 1 - d/\sigma)$ at $t = d/\sigma$.
Hence the residue theorem yields
\begin{equation}
\langle N_{\rm ex} \rangle \; \approx \; A \, \zeta(d/\sigma)
    \left(\frac{k_B T}{\Delta}\right)^{\!d/\sigma}  \; .
\label{EXOC}
\end{equation}
This canonical result, valid for $T < T_0$, coincides
precisely with the result of the customary grand canonical
analysis~[\ref{deGrootEtAl50}]. For example, in the case of the
three-dimensional isotropic harmonic oscillator potential (i.e.,
for $d = 3$, $\sigma = 1$, $A = 1$ and $\Delta = \hbar\omega$)
Eq.~(\ref{EXOC}) yields the familiar formula
\[
\langle N_0 \rangle \; = \; N - \langle N_{\rm ex} \rangle \; = \;
N\!\left[1 - \left(\frac{T}{T_0}\right)^3\right]
\quad {\mbox{for}} \quad T < T_0
= \frac{\hbar\omega}{k_B}\left(\frac{N}{\zeta(3)}\right)^{1/3} \; .
\]
\item If $d/\sigma = 1$, both Zeta functions in Eq.~(\ref{GREX})
coincide. We then encounter a double pole at $t = 1$, and find
\begin{equation}
\langle N_{\rm ex} \rangle \; \approx \; A \, \frac{k_B T}{\Delta}
    \left[ \ln\!\left(\frac{k_B T}{\Delta}\right) + \gamma \right] \; ,
\label{EXO2}
\end{equation}
where $\gamma = 0.5772\ldots$ is Euler's constant. This corresponds to a
result obtained already in 1950 by Nanda~[\ref{Nanda53}] with the help of
the Euler-Maclaurin summation formula.  
\item If $0 < d/\sigma < 1$, the pole of $\zeta(t)$ at $t = 1$ takes over:
\begin{equation}
\langle N_{\rm ex} \rangle \; \approx \;
    \frac{A}{\Gamma\!\left(\frac{d}{\sigma}\right)} \,
    \zeta(2-d/\sigma) \, \frac{k_B T}{\Delta} \; ,
\label{EXO3}
\end{equation}
so that for sufficiently low temperatures $\langle N_{\rm ex} \rangle$
now depends linearly on $T$, regardless of the value of $d/\sigma$ that
characterizes the trap.
\end{itemize}

\noindent
A mere glance at Eq.~(\ref{GRFL}) then suffices to reveal that
{\em the very same scenario\/} --- a first pole at $t = d/\sigma$ that
endows the temperature dependence with a trap-specific exponent as long
as it lies to the right of a second one, which yields universal behavior
when it becomes dominant --- also governs the canonical condensate
fluctuations, with the only difference that the second pole now is located
at $t = 2$:
\begin{itemize}
\item If $d/\sigma > 2$, the pole of $\zeta(t+1-d/\sigma)$ at
$t = d/\sigma$ wins, giving
\begin{equation}
\left(\delta N_0\right)^2_{cn} \; \approx \; A \, \zeta(d/\sigma-1)
    \left(\frac{k_B T}{\Delta}\right)^{\!d/\sigma}  \; .
\label{MFL1}
\end{equation}
\item If $d/\sigma = 2$, we find at $t = 2$ the already familiar double
pole, resulting in     
\begin{equation}
\left(\delta N_0\right)^2_{cn} \; \approx \; A \,
    \left(\frac{k_B T}{\Delta}\right)^2
    \left[ \ln\!\left(\frac{k_B T}{\Delta}\right) + \gamma + 1 \right] \; .
\label{MFL2}
\end{equation}
\item If $0 < d/\sigma < 2$, the pole of $\zeta(t-1)$ at $t = 2$ lies to
the right of its rival, yielding
\begin{equation}
\left(\delta N_0\right)^2_{cn} \; \approx \; 
    \frac{A}{\Gamma\!\left(\frac{d}{\sigma}\right)} \,
    \zeta(3-d/\sigma) \left(\frac{k_B T}{\Delta}\right)^2 \; .
\label{MFL3}
\end{equation}
In particular, for the one-dimensional harmonic oscillator we have
$d = 1$, $\sigma = 1$, $A = 1$ and $\Delta = \hbar\omega$, so that we
recover our previous microcanonical result~(\ref{ODOF}) within the
canonical ensemble, recalling that $\zeta(2) = \pi^2/6$. 
\end{itemize}
The above canonical fluctuations, derived from the integral
representation~(\ref{FUN2}), reduce to the expression obtained
by Politzer~[\ref{Politzer96}] in the case of the three-dimensional
isotropic trap, and match the results obtained by Wilkens and
Weiss~[\ref{WilkensWeiss97}].

The calculation of the corresponding {\em microcanonical\/} quantities  
now requires saddle-point inversions of Eq.~(\ref{RICH}) in order to
obtain the microcanonical moments $\mu_k(E)$ from the canonical
moments $M_k(\beta)$. Performing these inversions, and reexpressing
energy in terms of temperature, we find that the integral~(\ref{FUN1})
--- and, hence, the results~(\ref{EXOC}), (\ref{EXO2}), and (\ref{EXO3})
for the number of excited particles --- remains valid within the
microcanonical ensemble. The fluctuations require more care:
Whereas canonical and microcanonical fluctuations coincide in the
large-$N$-limit for $d/\sigma < 2$, the microcanonical mean-square
fluctuations $\left(\delta N_0\right)_{mc}^2$ are distinctly lower
than their canonical counterparts for $d/\sigma > 2$:
\begin{equation}
\left(\delta N_0\right)^2_{cn} - \left(\delta N_0\right)^2_{mc}
    \; \approx \; \frac{Ad}{d+\sigma}
        \frac{\zeta^2\!\left(\frac{d}{\sigma}\right)}
	     {\zeta\!\left(\frac{d}{\sigma}+1\right)}
	\left(\frac{k_B T}{\Delta}\right)^{\!d/\sigma}
	\quad {\mbox{for}} \;\; d/\sigma > 2 \;\;
	      {\mbox{and}} \;\; T < T_0 \; .
\label{DIFF}
\end{equation}
Thus, the exponent of $T$ is the same for both
$\left(\delta N_0\right)^2_{cn}$ and $\left(\delta N_0\right)^2_{mc}$,
but the prefactors can differ substantially. This Eq.~(\ref{DIFF})
contains as a special case the result obtained for the three-dimensional
isotropic trap by Navez {\em et al.\/}~[\ref{NavezEtAl97}].   

Before summarizing these findings, it is useful to also consider the heat
capacities for trapped ideal Bose gases with the single-particle
spectra~(\ref{SPSP}): for $d/\sigma > 1$, and temperatures below the
condensation temperature $T_0$, the heat capacity per particle is given by 
\begin{equation}
\frac{C_<}{Nk_B} \; = \; \frac{d}{\sigma}\!\left(\frac{d}{\sigma}+1\right)
    \frac{\zeta\!\left(\frac{d}{\sigma}+1\right)}
         {\zeta\!\left(\frac{d}{\sigma}\right)}
    \left(\frac{T}{T_0}\right)^{\!d/\sigma} \; ,
\end{equation}
above $T_0$ by 
\begin{equation}
\frac{C_>}{Nk_B} \; = \; \frac{d}{\sigma}\!\left(\frac{d}{\sigma}+1\right)
    \frac{g_{d/\sigma+1}(z)}{g_{d/\sigma}(z)}
    \, - \, \frac{d^2}{\sigma^2}\frac{g_{d/\sigma}(z)}{g_{d/\sigma-1}(z)} \; .
\end{equation}
Since the fugacity $z$ approaches unity from below when $T$ approaches
$T_0$ from above, so that the Bose function $g_\alpha(z)$ approaches
$\zeta(\alpha)$, we see that the heat capacity remains continuous at $T_0$
for $0 < d/\sigma \leq 2$, but exhibits a jump of size
\begin{equation}
\left.\frac{C_< - C_>}{Nk_B}\right|_{T_0} \; = \; \frac{d^2}{\sigma^2}
    \frac{\zeta\!\left(\frac{d}{\sigma}\right)}
         {\zeta\!\left(\frac{d}{\sigma}-1\right)}
\end{equation}
for $d/\sigma > 2$. 		

We thus arrive at the following picture: For any dimension~$d$ and trap
exponent~$\sigma >0$, the fluctuation of the number of condensate particles
is independent of the total particle number $N$. For isolated traps, this
insensitivity of the system with respect to $N$ reflects the well-peakedness
of the microcanonical distributions~(\ref{DIST}), see Fig.~2: if there is a
condensate, the behavior of the ideal Bose gas does not depend on how many
particles the condensate consists of. If $d/\sigma < 2$, so that the heat
capacity remains continuous in the large-$N$-limit, canonical and
microcanonical fluctuations $\delta N_0$ vanish linearly with temperature,
see Eq.~(\ref{MFL3}). If $d/\sigma = 2$, there appears a logarithmic
correction to the linear $T$-dependence, as quantified by Eq.~(\ref{MFL2}).
But if $d/\sigma > 2$, so that the heat capacity becomes discontinuous, then
the fluctuations $\delta N_0$ vanish proportionally to $T^{d/2\sigma}$,
so that now the properties of the trap determine the way the fluctuations
depend on temperature. In addition, in this case the microcanonical
fluctuations are markedly lower than the fluctuations in a trap that
exchanges energy with a heat bath.

Intuitively, one might have expected some sort of square root law for
the fluctuations. Because of the $N$-independence of the condensate
fluctuations, there is, of course, no ``$\sqrt{N}$-dependence'' of
$\delta N_0$. The square root is hidden elsewhere: Since
$\delta N_0 = \delta N_{\rm ex}$ for ensembles with fixed particle
number $N$, we find 
\begin{itemize}
\item
$\delta N_{\rm ex} \propto \langle N_{\rm ex} \rangle^{1/2} \quad \, $
for $ \quad 2 < d/\sigma$;
\item
$\delta N_{\rm ex} \propto \langle N_{\rm ex} \rangle^{\sigma/d} \quad $
for $ \quad 1 < d/\sigma < 2$; 
\item
$\delta N_{\rm ex} \propto \langle N_{\rm ex} \rangle \qquad \; $
for $ \quad 0 < d/\sigma < 1$,
\end{itemize}
with proportionality constants that are independent of temperature,
both canonically and microcanonically. 
The first of these relations is just what one might have guessed, but
the crossover from normal fluctuations for $d/\sigma > 2$ to much
stronger fluctuations for $0 < d/\sigma < 1$ appears noteworthy.

It remains to be seen how much of this ideal structure survives in the
case of weakly interacting Bose gases. It should also be recognized
that Maxwell's Demon, though it has provided the microcanonical
low-temperature fluctuations, can not solve all problems of the ideal
gas. When considering $d$-dimensional isotropic harmonic traps, the
Maxwell's Demon approximation (i.e., the replacement of the true upper
summation bound ``$N$'' in Eq.~(\ref{MIMO}) by ``$\infty$'') is
{\em exact\/} below the ``restriction temperature'' (i.e., that temperature
where the number $n = E/(\hbar\omega)$ of energy quanta equals the number
$N$ of particles~[\ref{GH97b}]), but the description of the Bose--Einstein
transition itself is beyond the capabilities of Maxwell's Demon. Namely,
that description requires the computation of the numbers
$\Omega(n\hbar\omega|N)$ of microstates also under conditions where the
restriction due to the finite~$N$ becomes decisive. Incidentially, one
meets the task of computing such restricted partitions of integers also
in other problems of statistical mechanics, for example in the theory of
the so-called compact lattice animals, or of the infinite-state Potts
model~[\ref{WuEtAl96}].

Nonetheless, the results obtained with the help of the Maxwell's Demon
approximation have some interesting number-theoretical implications.
Going once more back to Eq.~(\ref{ODOF}) for the one-dimensional
oscillator, and inserting the energy--temperature relation
$n = E/(\hbar\omega) \approx \zeta(2)(k_B T/\hbar\omega)^2$,
we find the truly remarkable formula
\begin{equation}
\delta N_0 \approx \sqrt{n} \; .
\label{AMAZ}
\end{equation}
This has a twofold interpretation. The physicist, puzzeled by the
loss of the square root fluctuation law at the level of
$\langle N_{\rm ex} \rangle $, finds a substitute:  
\begin{itemize}
\item
{\em For ideal Bose particles trapped at low temperatures by a
one-dimensional harmonic potential, the root-mean-square fluctuation
of the number of ground state particles is given by the square root of
the number of energy quanta.\/}
\end{itemize}
The mathematician, who approaches Eq.~(\ref{AMAZ}) from the viewpoint
of partition theory, sees the solution to another problem:
\begin{itemize}
\item
{\em If one considers all unrestricted partitions of the integer $n$
into positive, integer summands, and asks for the root-mean-square
fluctuation of the number of summands, then the answer is (asymptotically)
just $\sqrt{n}$\/}
\end{itemize}
--- certainly one of the most amazing examples for the occurrence
of square root fluctuations! The ease with which the solution to a
seemingly difficult number-theoretical question has been obtained
here is even aesthetically appealing. It is pleasing to conclude
that ongoing  developments in statistical mechanics, themselves
being motivated by recent experimental
achievements~[\ref{AndersonEtAl95},\ref{DavisEtAl95},\ref{BradleyEtAl97}],
have a high potential for further fertilization across subfield
boundaries.
\end{document}